\pdfoutput=1
\documentclass[
	physrev,
	reprint,
	amsmath,
	amsfonts,
	]{revtex4-2}
\usepackage{float}

\usepackage[all]{nowidow}
\usepackage{csquotes}

\usepackage{mathtools}

\usepackage[detect-all]{siunitx}
\usepackage{dcolumn}
\usepackage{booktabs}
\usepackage{color}

\begin{document}

\title{Relativistic diffusion model for hadron production in p-Pb collisions at the LHC}

\author{Philipp Schulz}
\author{Georg Wolschin}
\email{wolschin@thphys.uni-heidelberg.de}
\affiliation{Institute for Theoretical Physics, Heidelberg University, Philosophenweg 16, 69120 Heidelberg, Germany, EU}

\date{\today}

\begin{abstract}
We investigate charged-hadron production in relativistic heavy-ion collisions of asymmetric systems within a nonequilibrium-statistical framework.
Calculated centrality-dependent pseudorapidity distributions for p-Pb collisions at $\sqrt{s_{NN}}=5.02$ and 8.16 TeV are compared with data from the Large Hadron Collider (LHC).  Our approach combines a relativistic diffusion model with formulations based on quantum chromodynamics  {while utilizing} numerical solutions of a Fokker--Planck equation to account for the shift 
and  {broadening} of the  {fragmentation sources for particle-production} with respect to the stopping (net-baryon) rapidity distributions.
To represent the centrality dependence of charged-hadron production in asymmetric systems over  {a broad region} of pseudorapidities, 
the consideration and precise modelling of the fragmentation sources --  {along with the} central gluon-gluon source -- is found to be essential. 
 {Specifically}, this results in an inversion of the particle-production amplitude from backward- to forward-dominance when  {transitioning} from central to peripheral collisions, in agreement with recent ATLAS and ALICE p-Pb data  {at $\sqrt{s_{NN}}=\SI{5.02}{\tera\electronvolt}$}.

\end{abstract}

\maketitle

\section{\label{sec:introduction}Introduction}
In relativistic heavy-ion collisions  {involving} asymmetric systems such as d-Au at the BNL Relativistic Heavy Ion Collider  {(RHIC) or p-Pb} at the CERN Large Hadron Collider (LHC),
hadron production can be accounted for as occuring from three sources: a forward-going source arising from the leading particles and the interactions of  {their} partons with  {those} of the backward-going nucleus, a central-rapidity source \cite{Bjorken-1983-Phys.Rev.D27}  {mainly attributable} to gluon-gluon interactions, and a backward-going source caused by the excited fragment participants of the backward-going nucleus. The relative contributions of the three sources change substantially as function of centrality. In particular, the role of the fragmentation sources becomes more pronounced in peripheral collisions, where they can produce an inversion of the maximum of the particle-production amplitude from backward (Pb-going)  to forward (p-going). This effect has recently been observed experimentally by the ALICE collaboration in $\sqrt{s_{NN}}=5.02$ TeV p-Pb collisions at the LHC \cite{alice23}, and  {it calls} for a theoretical explanation.

In relativistic collisions of symmetric systems, the role of the fragmentation sources in particle production is less pronounced, but still relevant. In Pb-Pb or Au-Au, a spatially extended fireball is formed and provides the dominant central-rapidity source for particle production. The contribution of the fragmentation sources in symmetric systems had been discussed in a phenomenological three-sources diffusion model in rapidity space \cite{gw13,gw15}, using the proper Jacobian transformation for a mean transverse momentum to pseudorapidity ($\eta$) space. Agreement with centrality-dependent ALICE  {$\mathrm{d}N/\mathrm{d}\eta$} data has been achieved. We have recently extended the model to simultaneously treat transverse-momentum and rapidity variables \cite{hgw24}.
In this three-dimensional model with cylindrical symmetry, the contribution of the fragmentation sources was found to be smaller than in the  {one dimensional} model when compared to  {$\mathrm{d}^3N/(\mathrm{d}\eta\mathrm{d}p_{T}^2)$} ATLAS and ALICE data given the more stringent model constraints, but it remains non-negligible. 

Whereas in particle production, fragmentation and fireball sources cannot be disentangled experimentally, there is no central-rapidity source in stopping (net-baryon transport)
\cite{mtw09,hgw20,hhw23}, because particles and antiparticles are produced in equal amounts in the fireball and therefore, do not contribute to net-baryon distributions. Hence, the existing experimental net-proton (proton minus antiproton) data at SPS \cite{app99} and RHIC \cite{bea04} energies directly document the physical reality and significance of the fragmentation sources. This suggests that their role in hadron production cannot be neglected, although it is difficult to exactly pin down their contribution in symmetric systems.

In this work, we concentrate on a three-sources model for charged-hadron production in asymmetric systems, where it is easier to access the contribution of the fragmentation sources through the centrality-dependence of charged-hadron production than in symmetric systems. We focus on pseudorapidity distributions in p-Pb collisions at LHC energies of $\sqrt{s_{NN}}=5.02$ and 8.16 TeV, with special emphasis on the relative role of the three sources for particle production as function of centrality.

 We had already performed related, but more phenomenologically-oriented investigations of d-Au at  {$\sqrt{s_{NN}}=\SI{200}{\giga\electronvolt}$} \cite{wobi06} in comparison with PHOBOS data \cite{alver11}
 from RHIC, and p-Pb at 5.02 TeV \cite{sgw15,sgw18} in comparison with ALICE data \cite{adam15}.
These works are nonequilibrium-statistical model calculations without explicit connection to the underlying nonperturbative quantum-chromodynamical (QCD) physics. The initial rapidity distributions at $t=0$ were assumed to be delta-functions or Gaussians, thus providing an analytical solution of the time-dependent problem, and the time evolution for constant diffusion coefficients and linear drift was obtained from analytical solutions of a Fokker--Planck equation (FPE) in rapidity space. The transport parameters were determined in $\chi^2$-minimizations with respect to the available data, allowing for  predictions at higher energies through extrapolations of the parameters.

As a substantial refinement, we now aim to connect the relativistic diffusion model  (RDM) with elements from nonperturbative QCD. The initial states in the three sources are modeled as color-glass condensate (CGC) states, corresponding to valence-quark -- soft-gluon interactions as accounted for in the fragmentation sources for stopping, and gluon-gluon interactions for the central source. The subsequent nonequilibrium-statistical time evolution of the fragmentation sources is then calculated from a linear Fokker--Planck equation as before, but now numerical methods must be used for its solution because of the more involved initial conditions. The transport parameters governing the time evolution are determined in $\chi^2$-minimizations with respect to the available ATLAS and ALICE  {data of p-Pb collisions for charged-hadron production as functions of pseudorapidity and centrality $\mathrm{d}N/\mathrm{d}\eta$}. For the central gluon source, we take the color-glass distribution since the partial thermalization is less pronounced as compared to the fragmentation sources, where -- as it will turn out -- sizable drift and diffusion is actually observed. With this model, we aim to disentangle the respective role of the forward- and backward fragmentation sources, and the central-rapidity gluon source as functions of centrality in the full pseudorapidity range.

The CGC states for the central-rapidity source based on $k_{T}$-factorization, and the CGC initial states for the two fragmentation sources based on hybrid factorization are reviewed in the next section. The diffusion-model approach to the subsequent time evolution of the fragmentation distribution functions in rapidity space and the numerical solution of the corresponding FPE with the CGC initial conditions is considered in Sec.\,III. Results for charged-hadron production in pseudorapidity space for p-Pb collisions at $\sqrt{s_{NN}}=5.02$ and 8.16 TeV are presented in Sec.\,IV, and compared with ATLAS and ALICE data at various centralities. The role of the fragmentation sources is emphasized and the unexpected dominance of the proton-going source in very peripheral collisions is explained. Conclusions are given in Sec.\,V.

\section{CGC initial states}
Whereas in our earlier applications of the three-sources relativistic diffusion model to charged-hadron production in symmetric \cite{gw13,gw15} and asymmetric \cite{wobi06,sgw15,sgw18} systems, simplified initial conditions such as delta-functions or Gaussians were used to provide analytical solutions of the Fokker--Planck equation, we now determine the form of the initial state from a QCD-inspired model. For the fragmentation sources, this is analogous to the stationary state that is attained at the end of the baryon-transport (\textit{stopping}) process \cite{mtw09,hgw20,hhw23}. The central gluon source cancels out in stopping of symmetric systems, but is relevant in particle production.
We expect that most of the produced hadrons -- at least, in the fragmentation sources -- participate in the subsequent time-dependent thermalization process which is modelled through numerical solutions of the FPE for a constant diffusion coefficient as in our recent work on charged-hadron production in symmetric systems such as Pb-Pb \cite{hgw24}. We choose a linear dependence of the drift coefficient on the rapidity as in our earlier phenomenological results \cite{gw13,gw15,sgw18}, see the next section.

In addition to the two fragmentation sources, the model incorporates a central-rapidity source arising mostly from gluon collisions. In collisions of symmetric systems at RHIC and LHC energies, this is the dominant source for particle production from the hot fireball. For asymmetric collisions such as p-Pb, it is expected to be less pronounced, in particular, for more peripheral collisions. It is presently still a matter of discussion whether the mid-rapidity source (``QGP-droplet") exhibits collective phenomena similar to the ones in the spatially extended fireball that is formed in symmetric collisions such as Pb-Pb, but it certainly contributes to particle production. 

For the asymmetric p-Pb system, we take the p-going (\textit{projectile}) direction to define positive rapidities, while the Pb-going (\textit{target}) direction corresponds to negative rapidities. If necessary, data are mirrored to agree with this convention. The theoretical results as calculated in the center-of-mass system are converted to the laboratory system when comparing to data, taking the centrality- and energy-independent rapidity shift of $\Delta y = (1/2)\ln [Z_1\,A_2/(Z_2\,A_1)]$ with $\Delta y=0.465$ for p-Pb between the two systems in the direction of the proton beam into account. Charged-hadron production from the forward- and the backward-going fragmentation sources are calculated independently, and finally -- due to the linearity of the FPE -- added incoherently to the contribution of the central source.

For the initial particle distributions, we use a QCD-inspired framework  \cite{Gribov1983PR100,MuellerQiu1986NPB268,BlaizotMueller1987NPB289,McLerranVenugopalan1994PRD49} that is based on gluon saturation in the scattering of participant partons which we have already employed in our previous studies on baryon stopping \cite{mtw09,hgw20,hhw23}, and on charged-hadron production in central Pb-Pb collisions at LHC energies \cite{hgw24}. 
\subsection{Central gluon-gluon source}
Charged-hadron production from the central gluon-gluon source in relativistic collisions has commonly been described  {by} using $k_{T}$-factorization. The inclusive  {cross-section} \cite{kt02} must then be adapted to asymmetric collisions with distinct forward and backward processes. The result
\begin{equation}\label{eq:cross_section_N_gg}
\begin{split}
\frac{\mathrm{d}^3N^h_{gg}}{\mathrm{d}y \mathrm{d}{p}_T^2} = 
\frac{2}{C_F}
\frac{\alpha_s({m}_T^2)}{{m}_T^2}\\
\times\int_0^{p_T}
\mathrm{d}{k}_T^2
\left[\vphantom{\frac{1}{1}}\right. 
& N_{1}\varphi_{1}
\left(x_1,{k}^2_T\vphantom{\left|\right|^2}\right)
\varphi_{2}
\left(x_2, \left|{\textit{\textbf{p}}}_T-\textit{\textbf{k}}_T\right|^2 \right)\\
+& N_{2}
\varphi_{2}\left(x_2,{k}^2_T\vphantom{\left|\right|^2} \right)
\varphi_{1}
\left(x_1, \left|{\textit{\textbf{p}}}_T-\textit{\textbf{k}}_T\right|^2 \right)
\left.\vphantom{\frac{1}{1}}\right] 
\end{split}
\end{equation}
is a function of rapidity $y$ and transverse momentum ${p}_{T}$ of the produced hadron.
The unintegrated gluon distribution{s $\varphi_{1,2}(x,{k}_T^2)$ depend on the transverse momentum of the gluon $k_T^2$ and Bjorken-$x$, with $x_1=(m_{T}/\sqrt{s_{NN}})\,e^y$ and $x_2=(m_{T}/\sqrt{s_{NN}})\,e^{-y}$, $\varphi_1$ referring to the gluons in the proton, and  $\varphi_2$ to the ones in lead, A=Pb=208.} The transverse mass is $m_{T}=\sqrt{m^2+p_{T}^2}$, the color factor $C_F = {N_c^2-1}/{2N_c}$
such that $C_F=4/3$ for three colors as in our following calculations. 
We  {initially} scale the cross-section by the number of participants $N_1,N_2$ to account for the centrality dependence, rather than considering impact-parameter-dependent unintegrated gluon distributions. {However, when comparing with data, discrepancies remain, which are likely due to the centrality-dependence of nuclear suppression. We shall later account for this effect by considering an impact-parameter dependence of the gluon saturation scale, $Q_s^2(x)\rightarrow Q_s^2(x,b)$.}

{In Eq.\,(\ref{eq:cross_section_N_gg}), the presence of a large intrinsic momentum scale $Q_s$ for gluon saturation is taken to justify the use of a perturbative QCD expansion for non-perturbative observables such as hadronic cross sections, including pions \cite{kt02}. {Here, the effective quasiparticle mass $m\ll Q_s$ serves as an infrared regulator \cite{jpb10}. We have checked that the mass-dependence of $dN/dy$ calculated from Eq.\,(\ref{eq:cross_section_N_gg}) is sufficiently weak, such that the perturbative expression can also be used for produced pions}.  The unintegrated gluon distributions adhere to the different gluon saturation scales in the proton, and in Pb, causing strong {initial-state nuclear suppression \cite{alba10}} at low $p_T$ on the p-going (forward) side.}

For $\varphi(x,k^2)$ we use the Kharzeev-Levin-Nardi  {(KLN)} model 
\cite{Kharzeev2001b,Kharzeev2001, Kharzeev2004b,Kharzeev2005a,Duraes2016} 
\begin{equation}\label{KLN_UGD}
\varphi_\mathrm{KLN}\left(x,k^2\right) :=
\begin{dcases}
\frac{2C_F}{3\pi^2\alpha_s(Q^2)},  & k^2\leq Q^2_s(x)  \\[8pt]
\frac{2C_F}{3\pi^2\alpha_s(Q^2)}\frac{Q_s^2(x)}{k^2}, & k^2>Q^2_s(x)\,,
\end{dcases}
\end{equation}
where $k^2$ defines the internal momentum transfer scale.
The gluon saturation scale is 
\begin{equation}
\label{qs}
Q_s^2(x)=A^{1/3}Q_0^2(x/x_0)^{-\lambda},
\end{equation}
in agreement with the CGC literature. 
Specific parameters in our p-Pb calculations for the forward-rapidity central source are  {$x_0=1$, $A=1$, and $\lambda = 0.288$}.
The latter value is taken from fits to deep-inelastic scattering e-p data from HERA \cite{Golec-Biernat1998}, where $Q_0^2x_0^\lambda\simeq 0.097$ GeV$^2$ was determined together with $\lambda$. 
{These $(\lambda,Q_0^2)$-values where within experimental uncertainties found to be consistent with the ones needed for stopping in A-A collisions at SPS and RHIC energies \cite{mtw09}, and
charged-hadron production in Pb-Pb collisions at LHC energies \cite{hgw24}. }

For the backward-rapidity central source in p-Pb, we {thus initially use A=208. However, comparing to recent LHC p-Pb data, this will turn out to overestimate the gluon saturation scale in the nucleus when calculated according to Eq.\,(\ref{qs}), as already indicated in Ref.\,\cite{lap13}. Therefore, we shall later use $Q_0^2$ as a free parameter to not only correct for the $A^{1/3}$ scaling of Eq.\,(\ref{qs}), but also to capture the centrality dependence of the gluon saturation scale.}

Since the applicability of this approach is limited to small values of $x$, 
the unintegrated gluon distributions are usually modified according to \cite{alb07}
\begin{equation}\label{eq:KLN_1_x}
\widehat{\varphi}\left(k,x\right) = \left(1-x\right)^4 \varphi\left(k,x\right)
\end{equation}
in order to avoid overcounting for large value of  $x$, thus adhering to quark counting rules. Accordingly, we have introduced this
modification into the subsequent calculations.

The running of the strong coupling is accounted for as in Ref.\,\cite{alb13}. We use the parametrization
\begin{equation} \label{eq:QCDrunning_para_k}
\alpha_s(k^2) = \frac{4\pi}{\beta\ln\left( {{4k^2/\Lambda^2_\mathrm{QCD}}} +\mu\right)}\,.
\end{equation}
where $\beta = 11-2/3 N_f = 9$ with $N_f=3, \Lambda_\mathrm{QCD} = \SI{0.241}{\giga\electronvolt}$, and
$\mu$ regulating the strong coupling for large dipole sizes. It is determined by the condition $\alpha_s(\infty) = 0.5$, resulting in $\mu = 16.322$.
A very similar but slightly different parametrization of the strong-coupling constant has been used in \cite{alb18}.

\subsection{Fragmentation sources}
The fragmentation sources for charged-hadron production are made of valence-quark -- soft-gluon interactions in the forward direction, and vice-versa in the backward direction.
In addition, asymmetric collisions require separate calculations for both fragmentation sources by interchanging the roles of the projectile and target, and then adding the contributions incoherently.
In the backward direction, we adjust the valence-quark distribution functions $f_{q/A}$ of the nucleus by scaling them with the number of participants $N_\mathrm{part}$
$(x_1\equiv x_\text{p})$ 
\begin{equation}
x_A f_{q/A} = N_\mathrm{part} ~ x_p f_{q/p}\,,
\end{equation}
which is relevant for the correct calculation of the centrality dependence, see Sec.\,IV.
We obtain the initial distributions \cite{KharzeevKovchegovTuchin-2004-Phys.Lett.B599,BaierMehtarTaniSchiff-2006-Nucl.Phys.A764,DumitruHayashigakiJalilianMarian-2006-Nucl.Phys.A765}
as in our
previous stopping calculations for symmetric systems within the CGC model \cite{mtw09,hgw20,hhw23}, but now consider the significant forward-backward difference.

For valence quark-gluon scattering \cite{alt11}, we express the CGC initial condition for single-inclusive hadron production at rapidity $y$ and transverse momentum $p_T$ in the fragmentation sources of asymmetric proton-nucleus scattering as
\begin{eqnarray}
\label{eq:cross_section_N_qg}
\frac{\mathrm{d}^3N^h_{qg}}{\mathrm{d}y \mathrm{d}^2 p_T } = 
\frac{K}{(2\pi)^2}
\frac{1}{m^2_T}\qquad\qquad\qquad\qquad\qquad\qquad   \\\nonumber
\times\int_{x_F}^{1}\frac{\mathrm{d}z}{z^2} ~ 
D_{h/q}\left(z,\mu_f^2\right)
x_1 f_{q/p}\left(x_1,Q_f^2\right)
\varphi\left(x_2,q_T^2\right),
\end{eqnarray}
where $x_1$ is the Bjorken-$x$ of the valence quark in the proton, $x_2$ the one of a soft gluon in the nucleus $A\equiv$\,Pb, and $h\equiv \pi$, K, p stands for the produced charged hadrons and their antiparticles. 
The quark distribution function in a proton is $f_{q/p}(x_1,Q^2_f)$ with the factorization scale $Q_f^2$, the gluon distribution function in the nucleus is  {$\varphi(x_2,q_T^2)$}. We distinguish the produced hadrons that are considered explictly in our calculations by their masses and their Feynman-$x_F$, which is defined as
\begin{equation}\label{eq:feynman_x}
x_F = xp_T/k_T,
\end{equation}
where $p_T$ is the transverse momentum of the produced hadron and $k_T$ the transverse momentum of the parton \cite{fey69}.
Using the methodology of \cite{mtwc09}, the fraction $z$ of quark energy carried by the produced hadrons is
\begin{equation}
z(x):= x_F/x,
\end{equation}
where $z(1) = x_F$ and $z(x_F) = 1$ are the boundary conditions.
The differential $\mathrm{d}z$ is expressed as 
\begin{equation}
\mathrm{d}x/x_F = -x^2/x_F^2\,\mathrm{d}z = -\mathrm{d}z/z^2\,.
\end{equation}
Since massive constituents contribute an additional portion to the transverse momentum, we define an effective transverse momentum
\begin{equation}
q_T := m_T/z = \sqrt{p_T^2 + m^2}/z.
\end{equation}
For an effective impact parameter $\left<b\right>$, this expression corresponds to the minimum-bias cross section, with a fragmentation function $D_{h/q}$ of  quarks into hadrons, using
$Q_f^2= p_T^2$ for the factorization scale, and also $\mu_f^2 =p_T^2$ for the factorization scale of the fragmentation function. 
%
%
The factor $K$ accounts for higher-order corrections and additional dynamical effects that are not considered within the hybrid framework \cite{go13}.
 {We set $K=1$ and include these effects into $Q_0^2$.}
To obtain the full result for the initial distribution of the fragmentation sources in our charged-hadron production calculation, the role of projectile and target must be interchanged.

\section{Relativistic diffusion model}

Whereas the CGC distribution functions for the quark-gluon and gluon-quark fragmentation sources already provide an excellent description \cite{mtw09,hgw20,hhw23} of the measured stopping distributions at SPS and RHIC energies, they do not properly account for produced hadrons: As initial-state functions, they cannot consider the subsequent partial thermalization process and hence, their maxima occur at rapidity values too close to the beam rapidity when compared to final-state data. At LHC energies, this turns out to be more than two units of rapidity too large for a proper reproduction of the measured charged-hadron yields in p-Pb collisions.

 In order to take time-dependent  partial thermalization through drift and diffusion into account, we apply the relativistic diffusion model to the initial CGC distribution functions.
We have recently given a rigorous derivation  of the relativistic diffusion model in a framework for Markovian stochastic processes \cite{hgw24}. This allows to calculate the time evolution of particle-number distribution functions from solutions of a Fokker--Planck equation with respect to transverse and longitudinal rapidity, which are then transformed to transverse-momentum and pseudorapidity space and compared to data. In our previous Ref.\,\cite{hgw24}, we have confined the model to central collisions of heavy symmetric systems using cylindrical symmetry. 

Here, we focus on the centrality-dependence of charged-hadron production in asymmetric systems, where the role of the fragmentation sources is substantially more pronounced  compared to the central source, and the centrality dependence reveals interesting effects. For a correct description of the centrality dependence, the time evolution of the fragmentation sources towards equilibrium  is essential. Integrating over the transverse coordinates, one obtains a one-dimensional Fokker--Planck equation for the time evolution of the distribution function $R\,(y,t)$ in rapidity space \cite{gw99}
\begin{equation}
\frac{\partial}{\partial t}R\,(y,t)=-\frac{\partial}{\partial y}[J(y)R\,(y,t)] + D_y\frac{\partial^2}{\partial y^2}R\,(y,t)\,.
\end{equation}
For a constant rapidity diffusion coefficient $D$ and
 the conjecture that the stationary solution must be equal to a Maxwell-J\"uttner equilibrium distribution, the drift term $J(y)$ can be derived as \cite{la02} 
\begin{equation}
J(y) = -\frac{m_T D}{T}\sinh(y)\,,
\label{drift}
\end{equation}
with the transverse mass $m_T=\sqrt{m^2+p_{T}^2}$, the proton mass $m\equiv m_{p}$ and the equilibrium temperature $T$.
In leading order, this assumes the form of a linear relaxation term  \cite{gw99}
\begin{equation}
J(y) = \frac{y_{\mathrm{eq}}-y}{\tau_\mathrm{y}}\,,
\end{equation}
with the rapidity relaxation time $\tau_\mathrm{y}$ that governs the speed of the approach to thermal equilibrium. It is reached
at $y_{\mathrm{eq}}$, which is equal to zero for symmetric systems, or can be calculated from energy-momentum conservation for asymmetric systems \cite{gw99}.

For constant diffusion and linear drift as in the Uhlenbeck-Ornstein case \cite{uo30}, the FPE can be solved analytically in closed form in case of simple $\delta$-function or gaussian initial conditions. This approach has been used to calculate and predict charged-hadron distributions in symmetric systems, but also in d-Au and p-Pb \cite{wobi06,sgw18}. We have shown \cite{fgw17}
that the differences to a numerical solution with the full drift term Eq.\,(\ref{drift}) are small at RHIC energies, but may become more pronounced at LHC energies. In the present work, we use the CGC distribution as initial condition together with constant diffusion and linear drift. With the more sophisticated initial conditions that are provided by the CGC stopping distribution functions, only numerical solutions are possible. A corresponding \tt{C++} \rm code using the finite-element method has been written for this purpose.

The equilibrium distribution that would be approached in the course of the system's time evolution is taken to be the Maxwell-J\"uttner distribution
\begin{align}
E\frac{\mathrm{d}^3N}{\mathrm{d}p^3} &\propto E\exp(-E/T)\\ \nonumber
 &= m_T \cosh(y)\exp(-m_T\cosh(y)/T)\,,
  \end{align}
  with the transverse mass $m_\text{T}$ of the produced particle.
Accordingly, we express the time-dependent longitudinal production as
\begin{equation}
E\frac{\mathrm{d}N}{\mathrm{d}y}(y,t) = c\int_m^\infty m^2_T \cosh(y)R\,(y,t)~\mathrm{d}m_T\,.
\end{equation}
The underlying assumption is that the production yield can be described as the incoherent superposition of three distinct sources \cite{gw13}
\begin{align}
\frac{\mathrm{d}N^\mathrm{ch}}{\mathrm{d}y}(\tau_\mathrm{int}) =\qquad\qquad\qquad\qquad \qquad\qquad\qquad\qquad  \\ \nonumber
N_\mathrm{ch}^1 R_1\left(y,\tau_{\mathrm{int}}\right) +
N_\mathrm{ch}^2 R_2\left(y,\tau_{\mathrm{int}}\right) +
N_\mathrm{ch}^{gg} R_{gg}\left(y,\tau_{\mathrm{int}}\right),
\end{align}
where $N^{1,2}_\mathrm{ch}$ correspond to the produced charged hadrons in the fragmentation sources, while $N^{gg}_\mathrm{ch}$ accounts for charged-hadron production in the mid-rapidity gluon source.
The interaction or freezeout time $t=\tau_{\mathrm{int}}$ is the time span between first nuclear contact until separation, it is characteristic for the incomplete thermalization, and will be determined from the comparison of the calculated nonequilibrium distribution with the available data.

In asymmetric collisions, the mean value of the equilibrium rapidity $y_\mathrm{eq}(b)$ differs from zero \cite{gw13}.
When the beam rapidity $y_\mathrm{beam}$ is sufficiently large, it can be expressed as 
\begin{equation} \label{eq:yeq}
y_\mathrm{eq}(b)= \frac{1}{2}\ln\frac{\left<m^{(2)}_T(b)\right>}{\left<m^{(1)}_T(b)\right>}\,,
\end{equation}
with the average centrality-dependent transverse mass 
\begin{equation}
\left<m^{1,2}_T(b)\right> = \sqrt{m^2_{1,2}(b)+\left<p_T\right>^2},
\end{equation}
where $\left<p_T\right>$ is the average transverse momentum and $m^2_{1,2}(b) = m_p N_\mathrm{part}^{1,2}$ the participant mass. 
\begin{figure}[H]
	\includegraphics[width=\columnwidth]{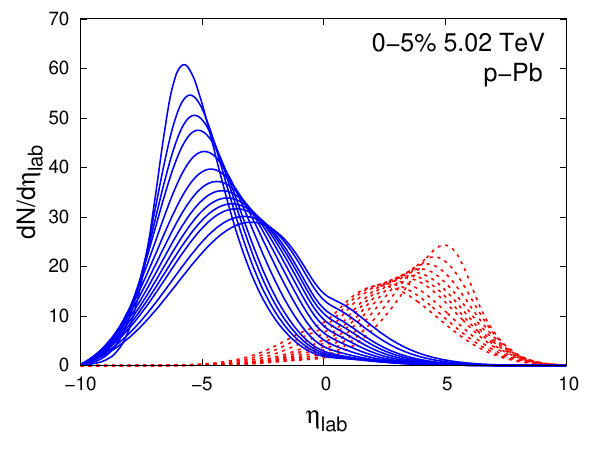}%
	\caption{\label{fig1}%
	Time-evolution of the quasiparticle fragmentation distributions as calculated in the relativistic diffusion model for 0-5\%
{p-Pb collisions} at $\sqrt{s_{NN}}=5.02$ TeV. The outermost initial distributions are calculated for baryon stopping within the hybrid-factorization scheme from valence-quark -- gluon interactions, whereas the final distributions correspond to produced charged hadrons. Various intermediate timesteps are shown for the Pb- and the p-going sides.  
(The midrapidity source is not shown here).}
\end{figure}
With the initial conditions derived from microscopic quark-gluon interactions for the fragmentation source, as well as for the mid-rapidity source obtained from gluon-gluon interactions,  we compute the corresponding rapidity distributions. Since data for charged-hadron distributions are obtained as functions of transverse momentum and pseudorapidity
$\eta=-\ln(\tan(\theta/2))$ with the scattering angle $\theta$, we transform \cite{gw13}  using the Jacobian transformation 
\begin{equation}\label{eq:d2NdetadpT}
\frac{\mathrm{d}^3N\left(\eta,p_T\right)}{\mathrm{d}\eta~\mathrm{d}p_T^2} =J(\eta,p_T)\frac{\mathrm{d}^3N\left(y,p_T\right)}{\mathrm{d}y~\mathrm{d}p_T^2}\,,
\end{equation}
\begin{equation}\label{eq:J_eff}
J(\eta,p_T) \equiv  \cosh(\eta)\left[1+(m/p_T)^2+\sinh^2(\eta)\right]^{-1/2}.
\end{equation}
For the gluonic source, we calculate the distribution as function of transverse momentum and rapidity with the above full Jacobian and subsequently perform a $p_T$-integration, whereas we use the transformation with an 
average $\langle p_T\rangle$ and an effective $\widehat{J}(m/\langle p_T\rangle)$ for the fragmentation sources, such that
\begin{equation}\label{eq:eff_jaocobian}
\frac{\mathrm{d}N}{\mathrm{d}\eta} = \widehat{J}\left( p \right)\frac{\mathrm{d}N\left(y\right)}{\mathrm{d}y}\,.
\end{equation}
A representative result for the pseudorapidity distributions solely from the fragmentation sources as function of time is shown in
Fig.\,\ref{fig1}:  The calculated diffusion process for the distribution functions originating from both fragmentation sources is displayed in a central (0-5\%) p-Pb collision at $\sqrt{s_{NN}}=5.02$ TeV. The parameters will be discussed in the next section.
Each fragmentation peak experiences a shift towards the equilibrium rapidity $y_\mathrm{eq}$, which is situated close to midrapidity. Concurrently, the distributions undergo a diffusion that enlarges their widths in rapidity space. The time interval is spanning from initial nuclear contact to the freezeout or interaction time $\tau_\mathrm{int}$. To obtain the full distribution function that is to be compared to data, the central gluon source will have to be added as described in the following section.
\section{Model results compared to LHC data}
To compare our model results with the available experimental  {$\mathrm{d}N/\mathrm{d}\eta$} centrality-dependent data for produced charge hadrons in p-Pb at $\sqrt{s_{NN}}= 5.02$ and 8.16 TeV from the ATLAS and ALICE collaborations, we integrate our results for the central source over $p_T$, add the diffusion-model result for the fragmentation sources that are obtained with an average $\langle p_T\rangle$, and perform a $\chi^2$-minimization to determine the values of the diffusion-model parameters. The CGC parameters of the initial state for the RDM time evolution  are kept fixed with $\lambda=0.288$, but to properly account for the centrality dependence, we allow for an impact-parameter dependent $Q_0\equiv Q_0(b)$.

For asymmetric collisions, distinct calculations are required for forward and backward rapidities. 
In case of p-Pb collisions at 5.02 TeV, the beam momenta are $\SI{4.0}{\tera\electronvolt}$ for the proton and $\SI{1.577}{\tera\electronvolt}$ for lead.
These beam momenta correspond to beam rapidities $y^p_\mathrm{beam} = 9.051$ and $y^{Pb}_\mathrm{beam} = 8.120$, respectively. 
The energy per nucleon-nucleon pair is thus $\sqrt{s_{NN}}=\SI{5.023}{\tera\electronvolt}$, which corresponds to a beam rapidity in the nucleon-nucleon frame of reference $y_\mathrm{beam} = 8.586$.
The rapidity shift between the laboratory frame and the nucleon-nucleon center-of-mass frame is  $\Delta y = 0.465$. 
In accordance with our convention, the ALICE data are transformed from the laboratory frame Pb-p to p-Pb. This involves transforming the experimental data to the center of mass frame, performing a mirroring operation, and then shifting it to the other laboratory frame.

\begin{figure}[H]
	\includegraphics[width=\columnwidth]{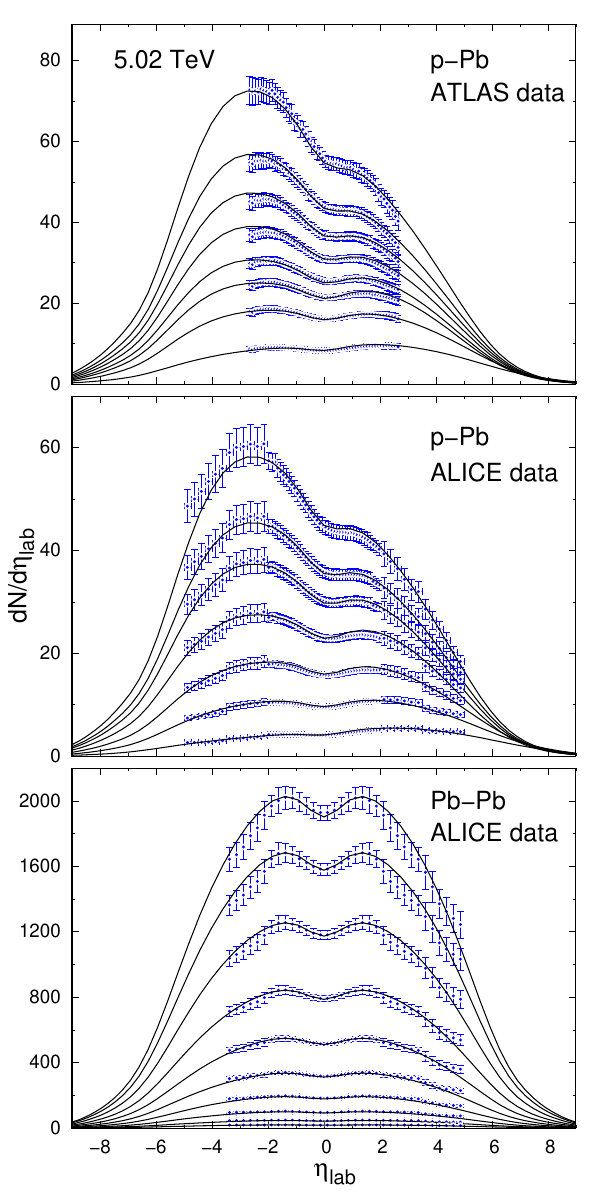}%
	\caption{\label{fig2}%
Calculated pseudorapidity distributions of produced charged hadrons (solid curves) in p-Pb {and Pb-Pb} collisions at $\sqrt{s_{NN}}= 5.02$ TeV compared with data for various centrality classes. Upper frame: Comparison with {p-Pb} data from ATLAS \cite{atlas16}. The centrality classes (from top to bottom) are 0-1\%, 1-5\%, 5-10\%, 10-20\%, 20-30\%, 30-40\%, 40-60\%, 60-90\%. Middle frame: Comparison with {p-Pb} data  from ALICE ($|\eta|< 2$ from \cite{adam15}, $|\eta| > 2.0$ from \cite{alice23}). The centrality classes are 0-5\%, 5-10\%, 10-20\%, 20-40\%,
40\%-60\%, 60-80\%, 80-100\%. Lower frame: Comparison with data from ALICE in Pb-Pb collisions at  $\sqrt{s_{NN}}= 5.02$ TeV in the centrality classes 0-5\%, 5-10\%, 10-20\%, 20-30\%, 30-40\%,
40\%-50\%, 50-60\%, 60-70\%, 70-80\% 80-90\% \cite{alice17}, with the same diffusion coefficient as on the Pb-going side in p-Pb.}
\end{figure}

The results for $\sqrt{s_{NN}}= 5.02$ TeV are shown in Fig.\,\ref{fig2}, with the ATLAS data from 2016 \cite{atlas16}  in eight centrality classes up to 90\% (upper frame), and the ALICE data from 2015 and 2023 in seven centrality classes plus minimum bias ($|\eta|<2$ from \cite{adam15}, $|\eta|>2$  from \cite{alice23}, middle frame). Parameters and results of the $\chi^2$-minimization results will be given in the next subsection. 
In the lower frame, we show for comparison a corresponding centrality-dependent calculation for Pb-Pb at $\sqrt{s_{NN}}= 5.02$ TeV, where we use the same diffusion coefficient as on the Pb-going side in p-Pb, thus achieving reasonable agreement with the ALICE Pb-Pb data \cite{alice17}.
\subsection{Charged-hadron distributions in ${{\sqrt{s_{NN}}=5.02}}$ TeV p-Pb}
\begin{figure*}[ht]
	\includegraphics[width=12cm]{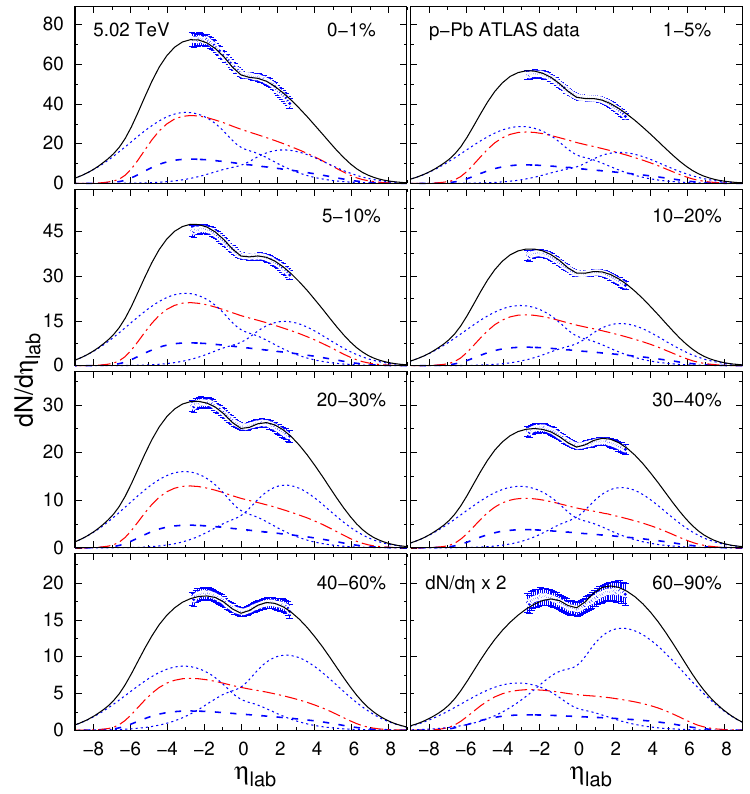}%
	\caption{\label{fig3}%
	Calculated pseudorapidity distributions of produced charged hadrons
in p-Pb collisions at $\sqrt{s_{NN}}= 5.02$ TeV for different centrality classes and $p_{T} > 0 $ (solid curves) compared with ATLAS data  \cite{atlas16}. The dot-dashed curves show the model contributions from the mid-rapidity source for produced pions, kaons and protons, whereas the dashed curves are only for pions. Kaons and protons are not explicitly displayed. The dotted curves represent the model contributions for pions coming from the two fragmentation sources,  {whereas the total yields also include kaons and protons as well as their antiparticles from the fragmentation sources. }}
\end{figure*}

We proceed to discuss charged-hadron distributions  {for p-Pb collisions at $\sqrt{s_{NN}}=5.02$ TeV} in more detail, presenting the individual contributions from both fragmentation sources, and the central source including their centrality dependence. The result is shown in a comparison with the ATLAS data \cite{atlas16} for eight centrality regions up to 60-90\% in Fig.\,\ref{fig3}.
The largest fraction of the charged-hadron yield is due to the gluon-gluon source at all centralities, but the relative contribution of the fragmentation sources -- which consist of charged pions, kaons and protons in our model calculation, and their antiparticles -- increases toward more peripheral collisions. 

Although this calculation does not extend to the most peripheral collisions, the model results clearly show that the amplitude of the forward-going fragmentation source becomes larger than the one of the backward-going source for more peripheral collisions. The consequence of this effect is already seen in the 60-90\% ATLAS data, although it is covered by the substantial asymmetry of the gluon-gluon source towards the Pb-going side. The effect will, however, more clearly be displayed when comparing to the more recent {very} peripheral 80-100\% ALICE data. 
The origin of this new effect -- which is obviously not present in peripheral collisions of symmetric systems --  is the strong gluon field in the Pb nucleus that the valence quarks in the forward-going proton experience.

The yields of produced charged hadrons ($\pi$, K, p) in the three sources are summarized in Tab.\,\ref{tab1}. The number of produced charged hadrons falls monotonically towards peripheral collisions, with central collisions yielding approximately six times more hadrons compared to  {the most} peripheral collisions. The ratio $R_p^{Pb}$ gives the particle yield produced in the Pb-going relative to the one from the p-going fragmentation source, and $R^{gg}_{qg}$ is the proportion of charged hadrons originating from the mid-rapidity source relative to the one from both fragmentation sources. It has a scaling  {behavior of} 
\begin{equation}
R^{gg}_{qg} \sim \frac{A^{1/3} N_\mathrm{part}}{A^{1/3}+N_\mathrm{part}}\,.
\end{equation}
The diffusion-model parameters and the average numbers of participants for each centrality window are listed in Tab.\,\ref{tab2}. 
The numbers of participants $N_\mathrm{part}^{Pb}$ are obtained from the ATLAS collaboration, where they were  calculated using Glauber Monte Carlo simulations \cite{atlas16}.
The respective interaction times $\tau_\text{in}^{p,Pb}$ on the p- and the Pb-going side are given with respect to the rapidity relaxation time $\tau_y$, thus avoiding the determination of an absolute timescale, which is not an observable. Values of the relative time scales are $t_p/\tau_y = 0.6, t_{Pb}/\tau_y = 0.6$ and the diffusion coefficient is $D^{Pb} = 12\tau_y^{-1}$. We use $\langle p_T\rangle=0.5$ GeV and $\langle m \rangle=m_\pi$  in the calculation of the Jacobian. Results from minimizations of our model results with respect to the ATLAS data at each centrality show reasonable $\chi^2/N_\text{dof}$ values.

\begin{figure*}[ht]
	\includegraphics[width=12cm]{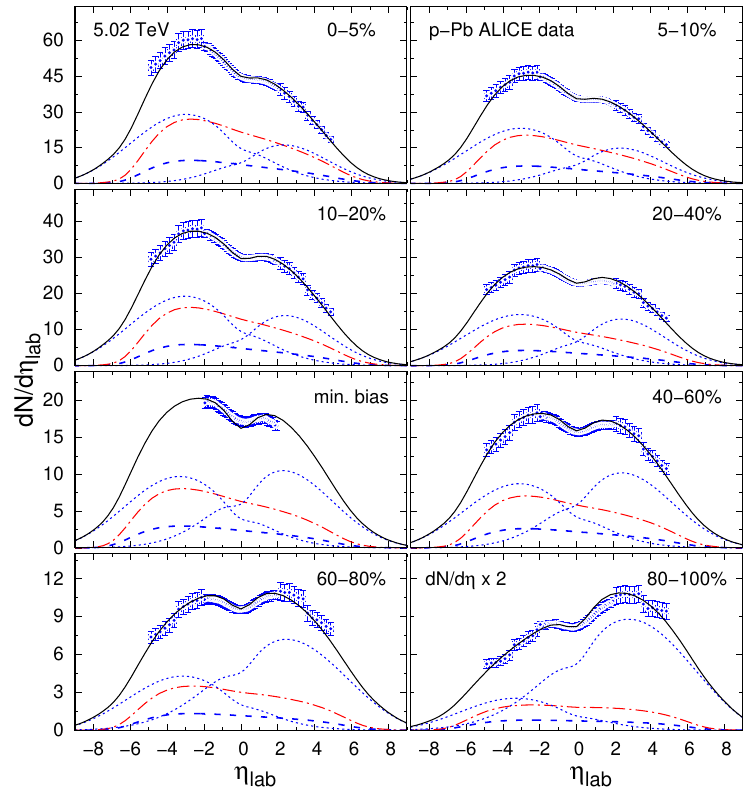}%
	\caption{\label{fig4}%
	Calculated pseudorapidity distributions of produced charged hadrons
in p-Pb collisions at $\sqrt{s_{NN}}= 5.02$ TeV compared with centrality-dependent ALICE data.
The minimum-bias data come from \cite{alice13}.
The data for the different centrality classes and $|\eta| < 2.0$ are from \cite{adam15} , whereas for $|\eta| > 2.0$ from \cite{alice23}. The solid distributions show the calculated pseudorapidity distributions, whereas dot-dashed curves are the combined distribution of charged hadrons produced by $gg$ interactions for pions, kaons and protons. The dashed distributions represent only the pions produced by $gg$ interactions and the dotted distributions are produced pions coming from $qg$ and $gq$ interactions  in the fragmentation sources. Kaons and protons are not shown explicitly for $gg$ and $gq$ interactions. The most peripheral distribution (80-100\%) is scaled by a factor of two for better visibility. Here, the amplitude on the p-going side becomes larger than the one on the Pb-going side due to the strong gluon field in the target. 
}
\end{figure*}

The results of our corresponding calculations in seven slightly different centrality classes and minimum bias compared to the ALICE p-Pb data at ${\sqrt{s_{NN}}=\SI{5.02}{\tera\electronvolt}}$ are displayed in Fig.\,\ref{fig4}. The earlier data set \cite{adam15} extends up to  {$|\eta_\text{lab}|\le2$, the later dataset is for $|\eta_\text{lab}|>2$} \cite{alice17}. In particular, the most peripheral region 80-100\% is included in these data, and with the new data set, a larger region in pseudorapidity space is covered as compared to the previous ATLAS data.
Regarding the ratio of produced particles in the midrapidity source relative to the one in the fragmentation sources shown in Tab.\,III, the results from the analysis of the ATLAS data are confirmed. Diffusion-model parameters and $\chi^2$ results are given in Tab.\,IV. Values of the relative time scales are $t_p/\tau_y = 0.6$ (except for 80-100\%),  $t_{Pb}/\tau_y = 0.6$ and the diffusion coefficient is $D^{Pb} = 12\tau_y^{-1}$. We use $\langle p_T\rangle=0.5$ GeV and $\langle m \rangle=m_\pi$  in the calculation of the Jacobian.

As a significant outcome of this investigation, an effect that had already shown up in the comparison with the ATLAS data is now clearly confirmed: In {very} peripheral collisions 
(60-80\% and, in particular, 80-100\%), the maximum of the particle-production amplitude moves from the Pb-going (backward) to the p-going (forward) pseudorapidity region.
Although the midrapidity source still has a slight preference toward the backward region, the fragmentation source in the forward region becomes much stronger than the backward one, thus causing the observed particle-production dominance on the p-going side in {the most} peripheral collisions. The origin of this effect is the strong gluon field that the valence quarks  in the forward-going proton experience in the target, whereas the backward-going proton(s) in peripheral collisions feel only the relatively weak gluon field in a single proton.

With $R_p^{Pb}\simeq N_\text{part}^{Pb}A^{-1/3}$ the fraction of produced charged hadrons originating from both fragmentation sources, 
the p-going fragmentation source yields more hadrons than the Pb-going one for $A^{1/3}>N_\text{part}^{Pb}$, which is the case for peripheral collisions, see Tabs.\,{III,\,IV}. 

For {very} peripheral collisions, this is not compensated for by the asymmetry of the central source anymore (dot-dashed curve in Fig.\,\ref{fig4}), and therefore, it becomes visible in the data. As a consequence, model calculations that do not explicitly include the effect of the fragmentation sources may have difficulties to account for the data in peripheral collisions of asymmetric systems.

{The diffusion-model coefficients obtained by our comparisons with ATLAS 
(Tab.\,II) 
and ALICE  (Tab.\,IV) data at 5.02 TeV are consistent with each other, underlining their universality within the applied model. Small deviations occur because the mean numbers of participants from the ATLAS Glauber Monte Carlo calculations for different centrality classes are slightly different from the  ALICE results (e.g., 13.6 vs. 13.0 for 5-10\%), but the results for the transport coefficients are almost identical.}

\begin{table}[H]
\begin{center}
\begin{tabular} { c r r r r r r r c c }
\hline\midrule
cent.$(\%)$ & $N_\mathrm{ch}$ & $N_\mathrm{ch}^{gg}$ & $N_{\pi}^{gg}$ & $N_{K}^{gg}$ & $N_{p}^{gg}$ & $N_{\pi,Pb}^{qg}$ & $N_{\pi,p}^{qg}$ & $R^{Pb}_{p}$ & $R^{gg}_{qg}$ \\
\midrule
0-1 & 629 & 273.6 & 99.3 & 92.3 & 82.0 & 256.0 & 100.7 & 2.5 & 0.8 \\[1px] 
1-5 & 507 & 209.8 & 76.8 & 70.8 & 62.3 & 204.8 & 93.2 & 2.2 & 0.7 \\[1px] 
5-10 & 431 & 171.9 & 63.3 & 58.0 & 50.7 & 172.4 & 88.2 & 2.0 & 0.7 \\[1px] 
10-20 & 367 & 140.4 & 51.9 & 47.4 & 41.1 & 143.2 & 83.9 & 1.7 & 0.6 \\[1px] 
20-30 & 300 & 108.7 & 40.5 & 36.7 & 31.6 & 113.8 & 78.2 & 1.5 & 0.6 \\[1px] 
30-40 & 254 & 88.1 & 32.9 & 29.7 & 25.4 & 91.7 & 75.2 & 1.2 & 0.5 \\[1px] 
40-60 & 192 & 61.5 & 23.1 & 20.7 & 17.6 & 61.8 & 69.6 & 0.9 & 0.5 \\[1px] 
60-90 & 103 & 26.1 & 10.0 & 8.8 & 7.3 & 22.7 & 54.7 & 0.4 & 0.3 \\[1px] 
\hline\hline
\end{tabular}
\end{center}
\caption{Calculated produced charged hadrons $N_\mathrm{ch}$ in \mbox{p-Pb} collisions at $\sqrt{s_{NN}}=\SI{5.02}{\tera\electronvolt}$ for various centrality classes, determined with respect to ATLAS data \cite{atlas16}.
The calculations integrate over the full range of pseudorapidity.
The relativistic diffusion model comprises three individual production sources: the mid-rapidity source $N^{gg}_\mathrm{ch}$ and two fragmentation sources $N^{qg}_{p/Pb}$.
The produced charged hadrons for the mid-rapidity source show separately pions $N^{gg}_{\pi}$, kaons $N^{gg}_{K}$, and protons $N^{gg}_{p}$. 
}\label{tab1}
\end{table}
\begin{table}[H]
\begin{center}
\begin{tabular}{c c c c c c c c}
\hline\midrule
cent.$(\%)$ & $\left<N_\mathrm{part}^{Pb}\right>$ & $Q_0^2\left(\si{\giga\electronvolt}^2\right)$ & $D^p\tau_y$ & $\chi^2/N_\mathrm{dof}$ \\
\midrule
0-1  & 17.2 & 0.055 & 7.0 & 0.115 \\[1px]
1-5 & 15.1 & 0.047 & 7.0 & 0.115 \\[1px]
5-10 & 13.6 & 0.042 & 7.0 & 0.210 \\[1px]
10-20 & 12.0 & 0.038 & 7.0 & 0.338 \\[1px]
20-30 & 10.4 & 0.033 & 7.0 & 0.135 \\[1px]
30-40 & 8.8 & 0.030 & 7.0 & 0.147 \\[1px]
40-60 & 6.4 & 0.027 & 10.0 & 0.052 \\[1px]
60-90 & 3.0 & 0.018 & 15.0 & 0.161 \\[1px]
\hline\hline
\end{tabular}
\end{center}
\caption{\label{tab2} 
Model parameters of the relativistic diffusion model for calculated pseudorapidity distributions in \mbox{p-Pb} collisions at $\sqrt{s_{NN}}=\SI{5.02}{\tera\electronvolt}$,
optimized with respect to ATLAS data \cite{atlas16}.
The mean numbers of participants $\langle N^{Pb}_\mathrm{part}\rangle$ were taken from the ATLAS Glauber Monte Carlo calculations  for different centrality classes.
In asymmetric collisions participants for projectile and target are treated separately, where 
$\left< N_\text{part} \right> = N^{p}_\mathrm{part} + N^{Pb}_\mathrm{part}$ with $N^{p}_\mathrm{part}=1$.
}
\end{table}

\begin{table}[H]
\begin{center}
\begin{tabular} { c r r r r r r r c c }
\hline\midrule
cent.$(\%)$ & $N_\mathrm{ch}$ & $N_\mathrm{ch}^{gg}$ & $N_{\pi}^{gg}$ & $N_{K}^{gg}$ & $N_{p}^{gg}$ & $N_{\pi,Pb}^{qg}$ & $N_{\pi,p}^{qg}$ & $R^{Pb}_{p}$ & $R^{gg}_{qg}$ \\
\midrule
0-5 & 519 & 217.6 & 79.4 & 73.4 & 64.9 & 206.8 & 96.1 & 2.2 & 0.7 \\[1px] 
5-10 & 417 & 165.5 & 60.9 & 55.8 & 48.8 & 164.8 & 88.2 & 1.9 & 0.7 \\[1px] 
10-20 & 353 & 133.9 & 49.6 & 45.2 & 39.2 & 137.4 & 82.8 & 1.7 & 0.6 \\[1px] 
20-40 & 275 & 97.3 & 36.3 & 32.8 & 28.2 & 100.9 & 77.0 & 1.3 & 0.5 \\[1px] 
m.b. & 208 & 69.6 & 26.0 & 23.5 & 20.1 & 68.1 & 71.5 & 1.0 & 0.5 \\[1px] 
40-60 & 193 & 61.6 & 23.1 & 20.8 & 17.7 & 62.0 & 69.6 & 0.9 & 0.5 \\[1px] 
60-80 & 119 & 32.4 & 12.3 & 10.9 & 9.1 & 30.3 & 57.0 & 0.5 & 0.4 \\[1px] 
80-100 & 55 & 10.4 & 4.2 & 3.5 & 2.7 & 8.9 & 36.1 & 0.2 & 0.2 \\[1px] 
\hline\hline
\end{tabular}
\end{center}
\caption{
Calculated produced charged hadrons $N_\mathrm{ch}$ in \mbox{p-Pb} collisions at $\sqrt{s_{NN}}=\SI{5.02}{\tera\electronvolt}$ for various centrality classes and minimum bias,
determined with respect to the ALICE data \cite{adam15,alice17}.
Displayed are the mid-rapidity source $N^{gg}_\mathrm{ch}$ and two fragmentation sources $N^{qg}_{1,2}$.
The produced charged hadrons include pions, kaons, and protons.
whereas the fragmentation regions only contain pions.
The ratio between the two fragmentation sources are shown as $R_{p}^{Pb} = N^{Pb}/N^{p}$, 
whereas the ratio $R_{gg}^{qg} = N^{gg}/N^{qg}$ represents the proportion of produced charged hadrons in the mid-rapidity region compared to the combined fragmentation regions.
}\label{tab3}
\end{table}
\begin{table}[H]
\begin{center}
\begin{tabular}{c c c c c c c c}
\hline\midrule
cent.$(\%)$ & $\left<N_\mathrm{part}^{Pb}\right>$ & $Q_0^2\left(\si{\giga\electronvolt}^2\right)$ & $D^p\tau_y$ & $\chi^2/N_\mathrm{dof}$ \\
\midrule
0-5 &  14.70 & 0.050 & 7.0 & 0.296 \\[1px]
5-10 &  13.00 & 0.042 & 7.0 & 0.108 \\[1px]
10-20 &  11.70 & 0.037 & 7.0 & 0.078 \\[1px]
20-40 &  9.40 & 0.032 & 7.0 & 0.256 \\[1px]
m.b. &  6.90 & 0.029 & 10.0 & 0.652 \\[1px]
40-60 &  6.42 & 0.027 & 10.0 & 0.288 \\[1px]
60-80 &  3.81 & 0.020 & 15.0 & 0.272 \\[1px]
80-100 &  1.94 & 0.009 & 19.0 & 0.320 \\[1px]
\hline\hline
\end{tabular}
\end{center}
\caption{
Model parameters of the relativistic diffusion model for calculated pseudorapidity distributions in \mbox{p-Pb} collisions at $\sqrt{s_{NN}}=\SI{5.02}{\tera\electronvolt}$ determined from ALICE data \cite{adam15,alice23}.
The number of participants $N^{Pb}_\mathrm{part}$ were taken from Glauber Monte Carlo calculations by ALICE for different centrality classes.
In asymmetric collisions participants for projectile and target are treated separately, where 
$\left< N_\text{part} \right> = N^{p}_\mathrm{part} + N^{Pb}_\mathrm{part}$ with $N^{p}_\mathrm{part}=1$.
}\label{tab4} 
\end{table}

\subsection{Charged-hadron distributions in ${{\sqrt{s_{NN}}=8.16}}$ TeV p-Pb}

At the highest currently available energy at the LHC, we compare our calculation for the same centrality classes as before with data for charged-hadron production in ${{\sqrt{s_{NN}}=8.16}}$ TeV p-Pb from the ALICE collaboration, which are so far confined to a smaller pseudorapidity interval $\eta_\text{lab}<1.8$ \cite{alice19}.

The beam momenta are  {$\SI{6.5}{\tera\electronvolt}$ for the proton and $\SI{2.563}{\tera\electronvolt}$} for the lead beam. 
The corresponding beam rapidities are $y^p_\mathrm{beam} = 9.536$ and $y^{Pb}_\mathrm{beam} = 8.606$. 
This configuration results in an energy per nucleon-nucleon pair of $\sqrt{s_{NN}}=\SI{8.162}{\tera\electronvolt}$, with a beam rapidity in the nucleon-nucleon frame of reference of $y_\mathrm{beam} = 9.071$.
(The rapidity shift of $\Delta y= 0.465$ between the laboratory frame and the nucleon-nucleon pair reference frame is independent of energy).

\begin{figure*}[ht]
	\includegraphics[width=8.9cm]{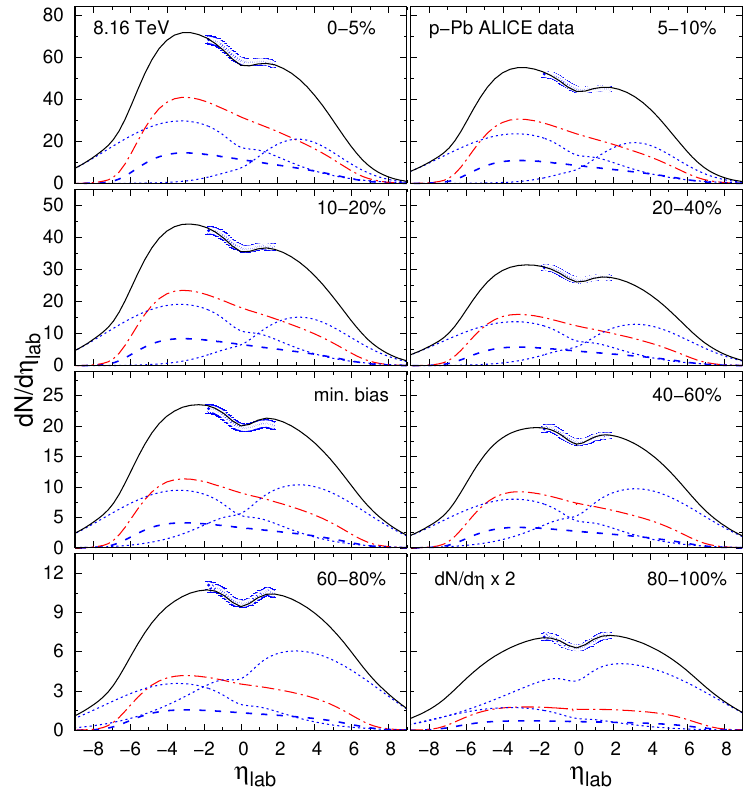}%
	\caption{\label{fig5}%
	Calculated pseudorapidity distributions of produced charged hadrons
in p-Pb collisions at $\sqrt{s_{NN}}= 8.16$ TeV compared with ALICE data \cite{alice19}
for different centrality classes. The solid curves are produced charged-hadron distributions calculated in the relativistic diffusion model. The dot-dashed curves show the model contributions from the mid-rapidity source for produced pions, kaons and protons, whereas the dashed curves are contributions only for pions. Kaons and protons are not explicitly shown. The dotted curves represent the model contributions for pions coming from the fragmentation sources. The scale of the most peripheral centrality (80-100\%) is enhanced by a factor of two for better visibility. The experimental data are mirrored from Pb-p to p-Pb convention.}
\end{figure*}

The centrality-dependent results are displayed in Fig.\,\ref{fig5} and are found in agreement with the data -- which are, however, not yet available in the most interesting pseudorapidity region where the inversion of the particle-production amplitude in peripheral collisions occurs, as is evident in the calculation for the 80-100\% centrality class.
The numbers of produced charged hadrons are given in Tab.\,V, the diffusion-model parameters and $\chi^2$-values in Tab.\,VI. Values of the relative time scales are $t_p/\tau_y = 0.4$ (except for 80-100\%), $t_{Pb}/\tau_y = 0.5$ and the diffusion coefficient is $D^{Pb} = 24\tau_y^{-1}$. We use $\langle p_T\rangle=0.5$ GeV and $\langle m \rangle=m_\pi$  in the calculation of the Jacobian.
\begin{table}[H]
\begin{center}
\begin{tabular} { c r r r r r r r c c }
\hline\midrule
cent.$(\%)$ & $N_\mathrm{ch}$ & $N_\mathrm{ch}^{gg}$ & $N_{\pi}^{gg}$ & $N_{K}^{gg}$ & $N_{p}^{gg}$ & $N_{\pi,Pb}^{qg}$ & $N_{\pi,p}^{qg}$ & $R^{Pb}_{p}$ & $R^{gg}_{qg}$ \\
\midrule
0-5 & 723 & 342 & 123 & 115.3 & 104.0 & 268.2 & 120.3 & 2.2 & 0.9 \\[1px] 
5-10 & 576 & 259 & 93.6 & 87.3 & 78.0 & 212.5 & 110.7 & 1.9 & 0.8 \\[1px] 
10-20 & 469 & 201 & 73.2 & 67.8 & 60.0 & 171.7 & 101.7 & 1.7 & 0.7 \\[1px] 
20-40 & 350 & 140 & 51.3 & 47.1 & 41.3 & 122.8 & 92.2 & 1.3 & 0.6 \\[1px] 
m.b.  & 267 & 102 & 37.3 & 34.3 & 30.0 & 84.8 & 86.2 & 1.0 & 0.6 \\[1px] 
40-60 & 232 & 83.7 & 31.0 & 28.2 & 24.5 & 71.5 & 80.7 & 0.9 & 0.5 \\[1px] 
60-80 & 131 & 40.7 & 15.3 & 13.7 & 11.7 & 31.8 & 60.5 & 0.5 & 0.4 \\[1px] 
80-100 & 45 & 10.0 & 4.0 & 3.3 & 2.6 & 7.6 & 28.8 & 0.3 & 0.3 \\[1px] 
\hline\hline
\end{tabular}
\end{center}
\caption{ {
Calculated produced charged hadrons $N_\mathrm{ch}$ for \mbox{p-Pb} collisions at ${\sqrt{s_{NN}}=\SI{8.16}{\tera\electronvolt}}$ for different centrality classes determined from ALICE data \cite{alice19}.
Three individual production sources are considered, the mid-rapidity source $N_{gg}$ and two fragmentation sources $N_{i}^{qg}$.}
The ratio between the fragmentation regions  {is $R_p^{Pb} = N^{Pb}_{qg}/N^{p}_{qg}$}. 
The ratio between the mid-rapidity region and both fragmentation regions is  {$R_{qg}^{gg} = N^{gg}/N^{qg}$}.
The produced charged hadrons for the mid-rapidity region display pions, kaons and protons. 
whereas the fragmentation regions show only pions.
\label{tab5} 
}
\end{table}
\begin{table}[H]
\begin{center}
\begin{tabular}{ c r c c c c }
\hline\midrule
cent.$(\%)$ & $\left<N_\mathrm{part}^{Pb}\right>$ & $Q_0^2\left(\si{\giga\electronvolt}^2\right)$ & $D^p\tau_y$ & $\chi^2/N_\mathrm{dof}$ \\
\midrule
0-5 & 16.00 & 0.058 & 7.0 & 0.896 \\[1px]
5-10 & 14.00 & 0.049 & 7.0 & 0.373 \\[1px]
10-20 & 12.40 & 0.042 & 11.0 & 0.532 \\[1px]
20-40 & 9.90 & 0.035 & 13.0 & 0.268 \\[1px]
m.b.  & 7.09 & 0.033 & 20.0 & 0.554 \\[1px]
40-60 & 6.47 & 0.029 & 20.0 & 0.143 \\[1px]
60-80 & 3.53 & 0.021 & 36.0 & 0.452 \\[1px]
80-100 & 1.76 & 0.007 & 48.0  & 0.080 \\[1px]
\hline\hline
\end{tabular}
\end{center}
\caption{Relativistic diffusion model parameters for calculated pseudorapidity distributions in \mbox{p-Pb} collisions at $\sqrt{s_{NN}}=\SI{8.16}{\tera\electronvolt}$ as determined from $\chi^2$ minimizations with respect to ALICE data  \cite{alice19}. The number of participants $\langle N^{Pb}_\mathrm{part}\rangle$ were taken from the ALICE Glauber Monte Carlo calculations for different centrality classes.
In asymmetric collisions participants are treated separately, where 
$\left< N_\text{part} \right> = N^{p}_\mathrm{part} + N^{Pb}_\mathrm{part}$ with $N^{p}_\mathrm{part}=1$.
\label{tab6}
}
\end{table}
As was already evident at 5.02 TeV, the effect of the Jacobian transformation from rapidity to pseudorapidity on the midrapidity source is very small, such that the midrapidity minimum that is seen in the data at all centralities must be attributed to the smallness of the fragmentation sources at midrapidity, although the dominant overall contribution to particle production arises from the gluon-gluon source. The number of produced charged hadrons depends monotonically on centrality, with central collisions generating  {about} 15 times more hadrons compared to {the most} peripheral collisions. 
\subsection{Centrality dependence of the initial saturation scale}

The initial gluon saturation-scale momentum $Q_s^2$ is defined by $Q_0^2$ and the exponent, for which we take the literature values from fitting HERA e-p data as  {$Q_0^2=0.09$ GeV$^2$} and $\lambda=0.288$ \cite{Golec-Biernat1998}. For heavy-ion collisions, however, a possible centrality-dependence of the gluon saturation scale becomes relevant. Sometimes,
the geometry of the collision is considered through the dipole cross section, mainly involving adjustments to the thickness function \cite{alb13,Rezaeian2013}.


The dependence of $Q^2_s$ on the mass number $A$ can be regarded as an approximation applicable primarily to very large collision systems. 
However, the behaviour of this dependence when the mass number decreases, or when only a limited number of participants are involved remains largely unexplored.
\begin{figure}[h]
	\includegraphics[width=\columnwidth]{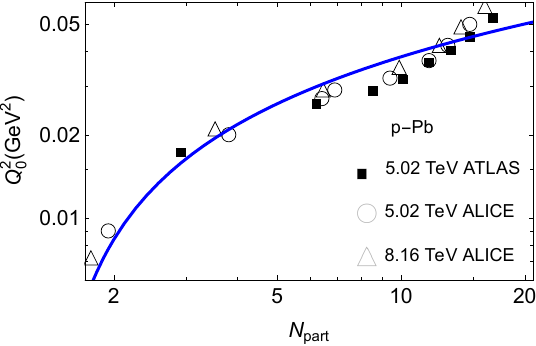}%
	\caption{\label{fig6}%
Centrality dependence of $Q_0^2$ in p-Pb at LHC energies as function of the number of target participants. The  {filled} squares refer to our theoretical results from $\chi^2$-minimizations with respect to ATLAS data at  $\sqrt{s_{NN}}= 5.02$ TeV \cite{atlas16}, while  {open} circles correspond to ALICE p-Pb data \cite{adam15,alice23}  at  $5.02$ TeV, and triangles to ALICE data  \cite{alice19} at $\sqrt{s_{NN}}= 8.16$ TeV. The solid curve is a two-parameter fit to all three data sets.	}
\end{figure}

In accordance with calculations presented in \cite{Dumitru2012}, we have adjusted
$Q_0^2$ depending on the number of participants  {$N_\text{part}$} in our centrality-dependent model calculations, as shown in Tabs.\,II,\,IV,\,VI. 
Using a simple two-parameter expression, $Q_0^2$  {as determined from the data} is found to rise strongly from central to peripheral p-Pb collisions at LHC energies by up to a factor of five, whereas the rise is much weaker in Pb-Pb at the same incident energy.
Fig.\,\ref{fig6} shows a double-logarithmic plot of the values of  {$Q_0^2(N_\text{part})$} that are required from $\chi^2$-minimizations of the calculated centrality-dependent pseudorapidity distributions in p-Pb collisions with respect to the ATLAS and ALICE data.

{By essentially considering $Q_0^2(N_\text{part})$ for the nucleus as a free parameter, we not only account for the centrality dependence beyond the mere participant-number scaling, but also implicitly correct the $A^{1/3}$ dependence of the gluon saturation scale in the nucleus: For} { minimum-bias collisions with a number of participants as calculated using Glauber Monte Carlo in Refs.\,\cite{atlas16,adam15}, }{ the value $Q_0^2\simeq 0.03$ as required from our fit to the centrality-dependent p-Pb data is about 1/3 of the HERA result for the proton. Even for the most central collisions, we obtain a significantly smaller value for $Q_0^2$ than what is expected from Eq.\,(\ref{qs}), thus offering a method for the determination of the gluon saturation scale in a nucleus.}

\section{Conclusions}
We have refined the three-sources relativistic diffusion model to include color-glass condensate initial conditions in an investigation of centrality-dependent p-Pb collisions at LHC energies of $\sqrt{s_{NN}}=5.02$ and 8.16 TeV. Whereas the largest contribution to charged-hadron production at all energies and centralities arises from the midrapidity gluon-gluon source, the relative contribution of the two fragmentation sources is found to be much more significant than in Pb-Pb collisions at LHC energies. Glauber-based calculations provided by the experimental collaborations are used for the numbers of participants in the centrality-dependent p-Pb calculations, with $N_\text{part}<18$.

Using the $k_T$- and hybrid-factorization schemes for small-$x$ gluon-gluon and valence-quark -- gluon interaction, respectively, we obtain the initial conditions, which we introduce into the relativistic diffusion model to account for the partial thermalization of the fragmentation sources in the time evolution of the collision. For both asymmetric fragmentation sources, the Fokker--Planck equation is solved numerically for constant diffusion coefficients and a linear dependence of the drift on rapidity.  Due to the linearity of the FPE, we can add the three sources incoherently. We perform Jacobian transformations to pseudorapidity distributions for charged hadrons produced at various centralities, and compare to recent data from the ATLAS and ALICE collaborations.

The computed pseudorapidity distributions accurately match the experimental p-Pb data from the ATLAS and ALICE collaborations at 5.02 TeV, as well as the one from ALICE at 8.16 TeV across a wide range of pseudorapidity values. The parameters of the initial-state CGC distributions are kept fixed except for an impact-parameter dependence of $Q_0^2$ that sets the scale for the gluon saturation momentum $Q_s^2$. The diffusion-model parameters are obtained in $\chi^2$-minimizations of the calculated $dN/d\eta$ distribution functions
with respect to the data, corresponding values at each energy and centrality are listed in the tables together with  {$\chi^2/N_\mathrm{dof}$}.

In particular, we have presented a first comparison with the new p-Pb ALICE data at 5.02 TeV in a large pseudorapidity range extending up to $\eta_\text{lab}=5$. The results include {very} peripheral events up to 100\% centrality, whereas ATLAS has provided results up to $\eta_\text{lab}\simeq 3$ and 90\% centrality. The significant role of the fragmentation sources becomes obvious in very peripheral collisions: Whereas in more central collisions, the maxima of charged-hadron production are on the Pb-going side due
to asymmetric gluon-gluon source that peaks in the backward region, in peripheral collisions the maxima shift towards the p-going side, as was already indicated in the early 60-90\% ATLAS data, and confirmed in the 80-100\% ALICE results.

This observation arises naturally in our three-sources relativistic diffusion model because the p-going fragmentation source in {peripheral} collisions is much larger than the Pb-going source due to the strong gluon field in the Pb nucleus that interacts with the few valence quarks in the forward direction. The effect is counteracted, but not overcome by the inherent bias of the central source towards the Pb-going side. The central source is often assumed to be the only one for particle production. In relativistic hydrodynamical calculations, it is unlikely that the observed amplitude inversion in particle production for peripheral collisions can be accounted for -- unless a three-fluid model is used.

At the higher energy of 8.16 TeV, our RDM-calculations exhibit similar trends, but the presently available data are still confined to the much smaller pseudorapidity region $|\eta_\text{lab}|<1.8$. As in case of the 5.02 TeV results, the dip observed
 around $\eta\!=\!0$ in charged-hadron production cannot be explained solely by the Jacobian transformation.
Instead, the smallness of the fragmentation sources in the central-rapidity region plays a significant role in the suppression around $\eta\!=\!0$, as these sources essentially peak at the same pseudorapidity where the distribution of charged-hadron production reaches its two local maxima.
This observation can be exploited to ascertain both the initial saturation scale and its dependence on centrality, which we have determined at both incident energies.

{The centrality-dependence of the gluon saturation scale in p-Pb has been accounted for using a participant-number dependence of the parameter $Q_0^2$. For minimum-bias collisions, the resulting saturation scale $Q_s^2$ is a factor of about three smaller than what is predicted by the usually assumed $A^{1/3}$ dependence, and for central collisions, it still remains significantly smaller, thus offering a method to determine the actual gluon saturation scale in a heavy nucleus.}

The model can be {used for predictions with suitably adapted transport coefficients. This may be of particular interest for the planned $\sqrt{s_{NN}}=$ { 9.62} TeV p-O pilot run in 2025. Our approach can also be}
 improved in various ways should future data in a larger pseudorapidity range become available to compare with. In particular, our simple assumption of a constant diffusion coefficient and a drift that depends linearly on the rapidity is easy to upgrade since we have solved the transport equation numerically already. The partial thermalization of the central source could also explicitly be considered, but would not change significantly the general outcome of the present investigation that puts the main emphasis on the transport properties of the fragmentation sources in the diffusion model once the initial conditions are prepared in the CGC model.

\begin{acknowledgments}
	We acknowledge discussions with Klaus Reygers about the ALICE p-Pb LHC data. GW is grateful to Farid Salazar for a discussion at LBNL Berkeley, {and to Alex Kovner at ITP Heidelberg.}
\end{acknowledgments}
%

\end{document}